\documentclass[twocolumn, pre]{revtex4}

\usepackage{amssymb}
\usepackage{amsmath}
\usepackage{graphics}
\usepackage{graphicx}
\usepackage{mathrsfs}
\usepackage{verbatim}

\newcommand{\be}{\begin{equation}}
\newcommand{\ee}{\end{equation}}

\begin{document}

\title{Anomalously large capacitance of an ionic liquid described by the restricted primitive model}

\date{\today}
\author{M. S. Loth}
\author{Brian Skinner}
\author{B. I. Shklovskii}
\affiliation{Fine Theoretical Physics Institute, University of Minnesota, Minneapolis, Minnesota 55455}

\begin{abstract}

We use Monte Carlo simulations to examine the simplest model of an ionic liquid, called the restricted primitive model, at a metal surface. We find that at moderately low temperatures the capacitance of the metal/ionic liquid interface is so large that the effective thickness of the electrostatic double-layer is up to 3 times smaller than the ion radius.  To interpret these results we suggest an approach which is based on the interaction between discrete ions and their image charges in the metal surface and which therefore goes beyond the mean-field approximation.  When a voltage is applied across the interface, the strong image attraction causes counterions to condense onto the metal surface to form compact ion-image dipoles.  These dipoles repel each other to form a correlated liquid.  When the surface density of these dipoles is low, the insertion of an additional dipole does not require much energy.  This leads to a large capacitance $C$ that decreases monotonically with voltage $V$, producing a ``bell-shaped" curve $C(V)$.  We also consider what happens when the electrode is made from a semi-metal rather than a perfect metal.  In this case, the finite screening radius of the electrode shifts the reflection plane for image charges to the interior of the electrode and we arrive at a ``camel-shaped" $C(V)$.  These predictions seem to be in qualitative agreement with experiment.

\end{abstract} \maketitle

\section{Introduction}  

Ionic liquids are molten salts made from ions which are large enough that their Coulomb interaction is relatively small, so that they remain in a fluid state at room temperature.  Essentially, an ionic liquid is a solvent-free electrolyte, which means that ionic liquids can be ideally suited for applications which require a thin or intensely concentrated layer of ionic charge.  Ionic liquids are already being used for batteries and ``supercapacitors" \cite{Galinski}, as well as for gating of new electronic materials.  It has therefore become a subject of great interest to understand the nature of the interface between an ionic liquid and a metallic electrode.

In its simplest form, the question of how an ionic liquid behaves in the vicinity of a charged metal surface seems remarkably straightforward.  While real-life experiments probing the structure of the ionic double-layer can be marked by a number of complications \cite{Bockris}, the essential description is encapsulated in a very simple model: an infinite, planar, metallic electrode is placed in contact with a semi-infinite volume with uniform dielectric constant $\varepsilon$ that contains a total concentration $N$ of mobile positively- and negatively- charged hard spheres, each with the same diameter $a$ and the same absolute value of charge $e$.  Such a model of the ionic liquid is called the ``restricted primitive model" (RPM).  If a voltage $V$ is applied between the electrode and the bulk of the ionic liquid, how large is the charge density $\sigma$ of the metal surface?  In other words, what is the capacitance per unit area $C(V) = d\sigma/dV$ of the interface?  


The answer to this question is well-known in the limit of low ion density, large temperature, and low applied voltage.  In this case the ionic double-layer is well-described as a diffuse screening layer with a characteristic size equal to the Debye-H\"{u}ckel (DH) screening radius
\be 
r_{DH} = \sqrt{ \frac{\varepsilon k_BT}{4 \pi e^2 N} }. \label{eq:rs}
\ee
Here, $k_BT$ is the thermal energy [Eq.\ (\ref{eq:rs}), and the remainder of this article, uses Gaussian units].  The diffuse layer of counter-charge effectively comprises the second half of a parallel-plate capacitor of thickness $r_{DH}$, so that the capacitance per unit area is equal to $C_{DH} = \varepsilon/4 \pi r_{DH}$.  This result for capacitance is valid as long as the ion density is low enough that $N a^3 \ll 1$, the temperature is high enough that $T \gg T_0 \equiv e^2/k_B \varepsilon a$, and the voltage is small enough that $eV \ll k_BT$.  Under this fairly extreme set of assumptions, the ionic double-layer is essentially a small perturbation of the bulk density $N$, so that the ion density and electric potential can be described using the linearized Poisson-Boltzmann equation.

More generally, one can characterize the capacitance by the effective thickness of the double-layer $d^* = \varepsilon /4 \pi C$.  In the DH limit, $d^* = r_{DH}$.  In realistic situations, however, the characteristic temperature $T_0$ is very large (for $a = 1$ nm and $\varepsilon = 3$, $T_0 \approx 5600$ K) and the DH approximation fails.  One can think that as a result ions become more strongly bound to the charged electrode and the size of the double-layer shrinks, so that $d^*$ decreases and the capacitance grows.  One may ask, then, how thin the double-layer can be, or in other words, how large the capacitance can be.  The apparent answer to this question goes back to Helmholtz \cite{Helmholtz1853}, who imagined that in an extreme case a neutralizing layer of ions could collapse completely onto the electrode surface, thereby forming the second half of a plane capacitor at a distance equal to the ion radius $a/2$.  The result is a double-layer of size $d^* = a/2$ and a capacitance per unit area equal to the ``Helmholtz capacitance"
\be 
C_H = \varepsilon/2 \pi a.
\ee 
In classical mean-field theories of the electrostatic double-layer \cite{Gouy1910, Chapman1913, Stern1924}, and in the recent influential theory of the metal/ionic liquid capacitance which accounts for the excluded volume among ions \cite{Kornyshev2007}, $C_H$ plays the role of a maximum possible capacitance per unit area.  Monte Carlo simulations \cite{Boda, Boda2, Kornyshev2008} seem to confirm this statement.  However, these and the majority of other simulations make the simplification of replacing the metal electrode by a uniformly-charged, insulating plane.  We argue below that in this way the essential physics of image charges in the metal surface is lost (see also Ref.\ \cite{PJ}, where some such simulations are critically analyzed).

It is the purpose of this article to demonstrate that capacitance $C > C_H$ is possible, or in other words, that the effective thickness of the double-layer can be smaller than the ion radius.  Our previous work \cite{us-longer} has demonstrated that capacitance $C > C_H$ can occur for highly asymmetric ionic liquids (where the cation has a much smaller radius than the anion, or vice-versa).  Here we show that even in the RPM, where cations and anions have equal diameter, capacitance significantly larger than the Helmholtz value is possible.  As we describe below, the metallic nature of the electrode --- specifically, the ability of ions to form image charges in the metal surface --- plays a key role in the development of large capacitance.  We present the results of Monte Carlo (MC) simulations of the restricted primitive model of an ionic liquid at various temperatures and densities, and we suggest a basic theoretical explanation of these results based on the weak repulsion between dipoles composed of bound ions and their images in the metal surface.

The remainder of this paper is organized as follows.  In section II we present our MC results for $C(T)$ at small voltage and give them a qualitative explanation. In section III we explain our MC procedure.  Section IV is devoted to our analytic theory,
including both temperature and voltage dependence of the capacitance $C(V, T)$.  Section V considers the role of the electrode material on $C(V,T)$.  Section VI discusses analogies between non-trivial capacitance phenomena at the metal/ionic liquid interface and those in semiconductor devices such as silicon MOSFETs and gated GaAlAs structures.  We conclude in section VII with a summary of our main results.

\section{Temperature dependence of the capacitance of the metal/ionic liquid interface}

\begin{figure}[htb]
\centering
\includegraphics[width=0.45 \textwidth]{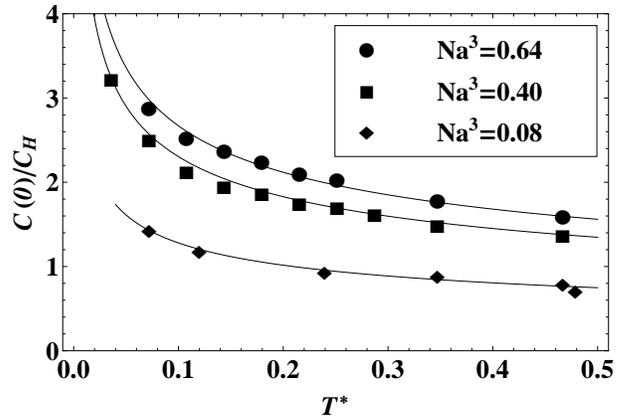}
\caption{The capacitance of the metal/ionic liquid double-layer at zero voltage as a function of the dimensionless temperature $T^*$, plotted for three values of the dimensionless ion density $Na^3$.  Symbols represent results from the MC simulation and solid lines show a best fit to the form $C/C_H = A \cdot (T^*)^{-1/3}$ for each density.  Error bars are smaller than the symbol size.} \label{fig:CT}
\end{figure}

Fig.\ \ref{fig:CT} shows the zero voltage capacitance $C(0)/C_H$, as measured by our MC simulations, as a function of reduced temperature $T^* = T/T_0$ for three different dimensionless ion densities $N a^3$.  The points correspond to results from the MC simulation, and solid lines are a fit to the form $C/C_H = A \cdot (T^*)^{-1/3}$, where $A$ is a numerical constant.  The motivation for this $(T^*)^{-1/3}$ dependence is explained in section IV.  For all three values of the density that we examined, the capacitance at low temperature is significantly higher than the Helmholtz value.  For practical applications, $Na^3 \approx 0.5$, and for $\varepsilon = 3$ and $a = 1$ nm room temperature corresponds to $T^* \approx 0.06$.  Using these parameters, the maximum value of capacitance in Fig.\ \ref{fig:CT} is $3 C_H \approx 27 \mu$F/cm$^2$.

These results should be contrasted with previous simulation studies \cite{Boda, Boda2} of the capacitance of the RPM ionic liquid, in which the metallic electrode was replaced by a charged insulator with uniform charge density $\sigma$. These studies report a capacitance $C(0)$ that grows with decreasing $T^*$ before reaching a peak at $T^* = T^*_p$ and then collapsing rapidly at $T^* < T^*_p$. For $Na^3=0.08$ and $0.64$, $T^*_p \approx 0.17$ and $0.28$, respectively.

The collapse of the capacitance at low temperatures $T^*
< T^*_p$ was interpreted by the authors of Refs.\ \cite{Boda, Boda2} as the result of strong binding of positive and negative ions to form neutral pairs. Such binding leads to an extreme sparsity of free charges in the ionic liquid, so that their total concentration $N_{f} \ll N$. Substituting $N_{f}$ for $N$ into Eq.\ (\ref{eq:rs}) at $T^* \ll 1$, we arrive at a large screening radius $r_{DH}$ and therefore much smaller capacitance $C(0)$.  These arguments are generic and convincing.  Why, then, does the capacitance in Fig.\ \ref{fig:CT} continue to grow with decreasing temperature?

Here we present a qualitative answer to this question. We begin
by observing that when the electrode is metallic, the energy of an ion binding to its image charge, $-e^2/2 \varepsilon a$, is exactly half the energy of a bound ion pair, $-e^2/\varepsilon a$. This fact implies that if an ion pair is separated in the bulk and then both ions are brought to the metal surface there is no net change in electrostatic energy. Thus, even in the absence of applied voltage there are plenty of free charges at the metallic surface. This allows the double-layer to be very thin and leads to the large capacitance shown in Fig.\ \ref{fig:CT}.

We note that the critical role of image charges for the structure of the double-layer has in fact been recognized by previous authors \cite{Bhuiyan}, who performed similar simulations which account explicitly for the electronic polarization of the electrode.  However, Ref.\ \cite{Bhuiyan} explored only very low ion density $Na^3 = 0.01$, where $C(0) < C_H$.  Another paper \cite{Madden} studied the capacitance of an ionic liquid between two identical metal plates and obtained large capacitance $C(0) \sim 2 C_H$, but this study used a much more complicated model of the ionic liquid.  

The following section explains our Monte Carlo procedure.

\section{Monte Carlo simulation} 

In our MC simulations, a canonical ensemble of $M_a$ anions and $M_c$ cations is placed in a square prism cell of dimensions $L \times L \times L/2$ and corresponding volume $\Omega = L^3/2$.  The metallic electrode coincides with one of the cell's square faces.  Every charge within the cell forms an electrostatic image in the electrode surface ($z = 0$), \textit{i.e.} a charge $q = \pm e$ at position $(x,y,z)$ has an image charge $-q$ located at $(x,y,-z)$.  The total electrostatic energy $\mathcal{E}$ of the cell is calculated as $1/2$ times the energy of a system twice as large composed of the real charges and their images, so that
\be 
\mathcal{E} = \frac12 \sum_{\{i,j\}}^{M_t} u(d_{i,j}), \label{eq:simE}
\ee 
where $M_t = 2(M_a +M_c)$ is the total number of charges in the system (ions plus images), $d_{i,j}$ is the distance between particles $i$ and $j$, and the two-particle interaction energy $u(d_{i,j})$ is
\be 
u(d_{i,j}) = \begin{cases} 
\infty, & d_{i,j} < a \\
q_i q_j /\varepsilon d_{ij}, & d_{i,j} > a
\end{cases}.
\ee
Here, $q_i = \pm e$ is the charge of ion $i$.

The charge of the electrode is varied by changing the number of anions $M_a$ and cations $M_c$ in the system by equal and opposite amounts, so that the total number of ions $M_a + M_c = N \Omega$ remains fixed for a given overall density $N$.  The corresponding electronic charge (in the form of image charges) in the electrode is $Q = e(M_a - M_c)$ and the capacitance $dQ/dV$ can be determined from the resulting voltage.  We use the system size $L = 20a$ everywhere.

At the beginning of each simulation, positive and negative ions are placed within the simulation cell in such a way that they do not overlap with each other or with the metal surface.  The MC program then selects an ion at random and attempts to reposition it to a random position within a cubic volume of $(2a)^3$ centered on the ion's current position.  The change in the energy $\mathcal{E}$ associated with this move is then calculated, and the move is accepted or rejected based on the standard Metropolis algorithm.  For one in every 100 attempted moves, the MC program chooses the random position from within a larger volume $(10a)^3$ as a means of overcoming the effects of any large, local energy barriers.  The simulation cell is given periodic boundaries, so that an ion exiting one face of the cell re-enters at the opposite face.  To ensure thermalization, 2,500 moves per ion are attempted before any simulation data is collected.  After thermalization, simulations attempt $2 \times 10^4$ moves per ion, of which $15\%$ -- $50\%$ are accepted.

The voltage of the electrode is measured by defining a ``measurement volume" near the back of the simulation cell --- occupying the range $-L/4 < x < L/4$, $-L/4 < y < L/4$, $L/4 < z < 3L/8$, where the origin $(x,y,z) = (0,0,0)$ is located at the center of the electrode surface --- inside of which the electric potential is measured.  After performing thermalization, the total electric potential is measured at 500 equally-spaced points within the measurement volume after every $3(M_a + M_c)$ attempted moves.  These measured values of potential are then averaged both temporally and spatially to produce a value for the voltage $V$ of the electrode relative to the bulk.  There was no noticeable systematic variation in electric potential across the measurement volume.  The  capacitance $C(V=0)$ is determined from the discrete derivative $\Delta Q/\Delta V$ at sufficiently small values of $Q$ for which the relationship $Q$ vs.\ $V$ is linear.  

Our results are shown in Fig.\ \ref{fig:CT}.  For all $Na^3$ studied, the lowest value of $T^*$ in Fig.\ \ref{fig:CT} is larger than the corresponding liquid-gas or liquid-solid coexistence temperature.  For comparison, the triple point in the phase diagram of the RPM ionic liquid is located at $Na^3 = 0.5$ and $T^* = 0.025$ while the gas-liquid critical point is at $Na^3 = 0.02$ and $T^* = 0.05$ \cite{Levin}.  We verified for each simulation that there was no phase separation within the simulation cell.  

In order to quantify the finite-size effects of our simulation cell, we examined the capacitance at zero voltage, $C(0, T^*)$, obtained from three simulation volumes of size $L = 10a$, $20a$, and $30a$.  For $Na^3=0.4$, $C(0, T^*)$ was seen to scale linearly with $1/L$ at all values of the temperature that we examined ($T^* = $ 0.042, 0.072, and 0.14).  In each case, the value of $C(0, T^*)$ obtained by extrapolation to infinite system size was within $20\%$ of the value of $C(0, T^*)$ corresponding to $L = 20a$.  These results allow us to conclude that the simulation cell with $L = 20 a$ provides a reasonable approximation of an infinite system. All MC results presented below correspond to this choice.

\section{Semi-quantitative theory of the capacitance of metallic electrodes}  

Our goal is to explain the large capacitance of the
metal/ionic liquid interface at $T^* \ll 1$. As we emphasized
above, an ionic liquid next to a metallic electrode has a high degree of degeneracy because of the zero-energy process of ion pairs in the bulk dissociating and sticking to their images on the metal surface. As a result, at low temperatures $T^* \ll 1$ effectively all ions in the system are either paired in the bulk or bound to their images on the metal surface \cite{clusters}.  At zero applied voltage, equal numbers of positive and negative ions are bound to the metal surface. The area density $n_0$ of these ions can be estimated from the requirement that the chemical potential of pairs in the bulk be equal to the chemical potential of ions at the surface, which gives $\ln(1/Na^3) \simeq 2
\ln(1/n_0a^2)$, so that $n_0 \simeq \sqrt{N/a}$.

\begin{figure}[htb]
\centering
\includegraphics[width=0.3 \textwidth]{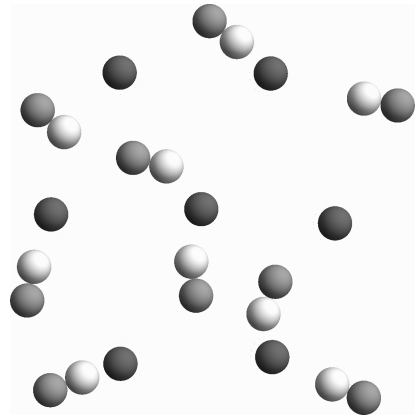}
\caption{A schematic depiction of the layer of ions bound to the metal surface at finite charge density $\sigma < 0$.  Cations (darkly-shaded spheres) and anions (lightly-shaded) are both bound to the metal surface by the strong attraction to their image charges.  Excess cations (made even darker for emphasis), which neutralize the electrode charge, minimize their repulsive energy by forming a Wigner crystal-like lattice.  All other ions remain bound in neutral pairs.} \label{fig:schematic}
\end{figure}

As the voltage $V$ of the electrode is increased from zero, some number of pairs in the system are separated so that the free counterion can come to neutralize the electrode surface.  The corresponding density of these ``excess ions" $\delta n$ on the metal surface is related to the charge density $\sigma$ by 
$\delta n = |\sigma|/e$. 
If $\sigma > 0$, then $\delta n$ represents an excess of anions on the surface; if $\sigma < 0$ the excess ions are cations.  Naturally, excess ions condensed onto the metal surface will repel each other.  Since each ion on the metal surface is separated by a distance $a$ from its image charge in the metal, ions and their images constitute compact ion-image dipoles with dipole moment $e a$, and so excess ions repel each other via a dipole-dipole interaction
\be 
u(\delta n) = \frac{e^2 a^2 (\delta n)^{3/2}}{2 \varepsilon}. \label{eq:udd}
\ee 
Excess ions at the metal surface are surrounded by $n_0$ other ions per unit area, which effectively neutralize each other by forming $n_0/2$ bound pairs (see Fig.\ \ref{fig:schematic}).  These $n_0/2$ bound pairs, along with bound pairs in the bulk, may serve to modify the effective dielectric constant for the interaction of excess ions.  We comment on this possibility later in this article.

At low temperatures, the excess ions will seek to maximize the distance from each other while maintaining a given density $\delta n$, which results in the formation of a strongly-correlated liquid of excess ions reminiscent of a two-dimensional Wigner crystal (see Fig.\ \ref{fig:schematic}).  The corresponding total electrostatic energy per unit area of the system is
\be 
U = \alpha \cdot \delta n \cdot u(\delta n) - \sigma V,
\ee 
where $\alpha$ is a numerical coefficient which describes the structure of the lattice of excess ions; for a triangular lattice, $\alpha \approx 4.4$ \cite{Topping}.  The term $- \sigma V$ describes the work done by the voltage source.  Fig.\ \ref{fig:schematic} shows a schematic depiction of the layer of ions bound to the metal surface.

The voltage $V$ which corresponds to a given charge density $\sigma$ can be found by the equilibrium condition 
$\partial U/\partial \sigma = \partial U/\partial (e \delta n) = 0$,
which gives
\be 
V = \frac{5 \alpha e a^2}{4 \varepsilon} (\delta n)^{3/2}. \label{eq:Vn}
\ee 
The resulting capacitance per unit area $C = d\sigma/dV = e [dV/d(\delta n)]^{-1}$ is
\be 
C(\delta n) = \frac{8 \varepsilon}{15 \alpha a \sqrt{\delta n a^2}}. \label{eq:Cn}
\ee 
Substituting Eq.\ (\ref{eq:Vn}) into Eq.\ (\ref{eq:Cn}) gives the capacitance in terms of voltage:
\be 
C(V) = \frac{8}{15} \left( \frac{5}{4 \alpha} \right)^{2/3} 
\left( \frac{e}{\varepsilon a V} \right)^{1/3} \frac{\varepsilon}{a}
\approx 1.4 \left( \frac{e}{\varepsilon a V} \right)^{1/3} C_H. \label{eq:CV}
\ee
This expression can be significantly larger than $C_H$ when $V$ is small.  Physically, at such small voltages the excess ions are very sparse, and so their mutual repulsion goes to zero.  In other words, at low voltages charging of the electrode is not limited by the accumulation of complete charged layers, but by the weak dipole-dipole repulsion between discrete, correlated ions and their image charges.  

Of course, the validity of Eqs.\ (\ref{eq:udd}) -- (\ref{eq:CV}) is limited to the range of voltage where there is a small fractional coverage of the metal surface by excess ions, $na^2 \ll 1$.  By Eq.\ (\ref{eq:Vn}), this corresponds to a dimensionless voltage $V^* = V/(e/\varepsilon a) \ll 5 \alpha /4 \approx 5.5$.  At large enough voltages that $na^2 \simeq 1$, excess ions constitute a uniform layer of charge, and therefore the capacitance approaches $C_H$.  At even larger voltages, the capacitance declines as complete layers of counterions accumulate next to the electrode and the double-layer becomes thicker.  This leads to a mean-field capacitance $C \propto V^{-1/2}$ at large voltages, as described in Ref.\ \cite{Kornyshev2007}.

Formally, Eq.\ (\ref{eq:CV}) diverges as the voltage goes to zero.  Of course, this expression neglects entropic effects of the excess ions, which tend to destroy the lattice structure of excess dipoles on the metal surface.  Such effects will truncate the low-voltage divergence of Eq.\ (\ref{eq:CV}), resulting in a finite capacitance at zero voltage.  The value of this capacitance maximum can be estimated roughly by setting $u(\delta n_c) = k_BT$, solving for the corresponding concentration $\delta n_c$, and then plugging $\delta n_c$ into $C(\delta n)$ from Eq.\ (\ref{eq:Cn}).  This gives
\be 
C_{max}(T) = \frac{8}{15 \sqrt[3]{2} \alpha} \left( \frac{e^2}{\varepsilon a k_B T} \right)^{1/3} \frac{\varepsilon}{a} 
= \frac{A}{(T^*)^{1/3}} C_H ,
\label{eq:CT}
\ee  
where $A \approx 0.6$.  In other words, the effective thickness $d^* = a (T^*)^{1/3}/2A$.  At $T^* \ll 1$, we find that $d^* \ll a$.  The corresponding voltage at which the capacitance plateaus ($u_{dd}$ becomes equal to $k_BT$) is $V^*_c = 5 \alpha T^*/2  \approx 11 T^*$.

\begin{figure}[htb]
\centering
\includegraphics[width=0.5 \textwidth]{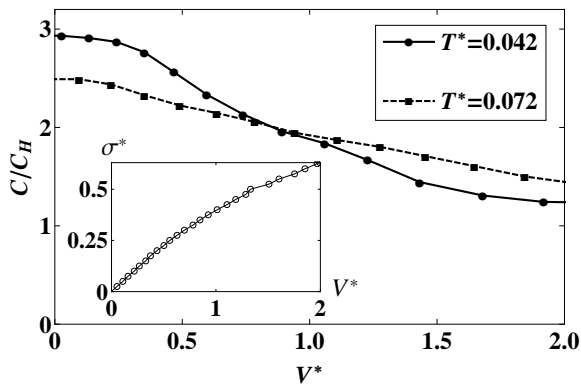}
\caption{The capacitance as a function of the dimensionless voltage $V^* = V/(e/\varepsilon a)$ at two different temperatures for a system with ion density $Na^3 = 0.4$.  The inset shows a plot of the dimensionless charge density $\sigma^* = \sigma a^2/e$ as a function of the voltage $V^*$ measured by the MC simulation at the temperature $T^* = 0.042$.  For $\varepsilon = 3$ and $a = 1$ nm, $V^* = 2$ corresponds to 0.96 Volts.} \label{fig:CV}
\end{figure}

Fig.\ \ref{fig:CV} shows the capacitance as a function of $V^*$, as measured by our MC simulation, at density $Na^3 = 0.4$ and at two values of the temperature $T^*$.  The inset shows the dimensionless charge density $\sigma^* = \sigma a^2/e$ of the electrode as a function of the voltage $V^*$ for the temperature $T^* = 0.042$.  The capacitance is determined by a numerical derivative of the $\sigma$ vs.\ $V$ curve.  Here, $e/a^2$ is the maximal density for a square lattice of ions on the metal surface, so that $\sigma^*$ can be interpreted as a filling factor of the first layer of ions.  Note that the capacitance drops substantially even at low filling factor $\sigma^*$, so that the capacitance is already reduced by a factor two at $\sigma^* = 0.5$.  This suggests that the decline in capacitance is not driven by the excluded volume effects emphasized in the theory of Ref.\ \cite{Kornyshev2007}.

The prediction of Eq.\ (\ref{eq:CT}) provides a good fit to the capacitance measured by our MC simulation at low ion density, as shown in Fig.\ \ref{fig:CT}.  However, while the dependence $C \propto (T^*)^{-1/3}$ remains accurate for all densities, the constant $A$ apparently depends on the ion density, taking the values $A = 0.6, 1.1$, and $1.2$ for $Na^3 = 0.08, 0.4$, and $0.64$, respectively.  This increase is also reflected in Fig.\ \ref{fig:CV}, where the capacitance at finite voltage is somewhat larger than predicted by Eq.\ (\ref{eq:CV}), consistent with the increase in the constant $A$.  This larger capacitance at high densities is perhaps an indication that the dipole interaction suggested in Eq.\ (\ref{eq:udd}) is weaker at large ion density.  One possible explanation is that at high densities ion pairs in the vicinity of two excess ions can polarize in the direction of the electric field, thereby producing an effectively larger dielectric constant.  If we replace $\varepsilon$ in Eq.\ (\ref{eq:udd}) by an effective dielectric constant $\varepsilon \varepsilon'$, then we find that $A \approx 0.6 (\varepsilon')^{2/3}$.  The values of the constant $A$ from above suggest that for bulk densities $Na^3 = 0.08, 0.4$, and $0.64$, the value of $\varepsilon'$ is $1.0, 2.4$, and $2.9$, respectively.  These values are consistent with our interpretation that the effective dielectric constant should increase with ion density, driving the capacitance upward.

Based on our arguments from this section about the dependence of the capacitance on voltage and temperature, we can hypothesize a general scaling relationship $C(V^*,T^*)$ which reproduces Eqs.\ (\ref{eq:CV}) and (\ref{eq:CT}):
\be 
\frac{C(V^*, T^*)}{C_H} = \frac{\beta_1}{\left[(\beta_2 T^*)^2 + (V^*)^2 \right]^{1/6}}. \label{eq:CVT}
\ee
Here, $\beta_1$ and $\beta_2$ are numerical coefficients.  Applying this fit to the curves shown in Fig.\ \ref{fig:CV} gives a reasonably good fit with $\beta_2 \approx 8$, suggesting that the capacitance plateaus at about $V^* = 8 T^*$, as compared to the theoretically estimated value $V^*_c = 11 T^*$.

\section{Electrode material: from perfect to poor metal} 

So far we have assumed that the electrode is a perfect metal, or
in other words, that the screening radius $b$ of the metal is much smaller than the ion diameter $a$. This assumption is justified for ionic liquids with large ions and electrodes made from a good metal. Experiments on such systems have indeed reported large capacitance that declines with absolute value of voltage \cite{Ohsaka} (the $C$--$V$ curve is ``bell-shaped"). However, for smaller ions and for electrodes made from semi-metals, such as graphite or glassy carbon, experimental values of $C(0)$ are smaller and the $C(V)$ curves are ``camel-shaped", \textit{i.e.} the capacitance grows parabolically near $V = 0$ ~\cite{Ohsaka,Ohsaka-lett,Lockett}.

In order to interpret this difference qualitatively, let us recall that in such poor metals the density of states at the Fermi level is relatively small and the screening radius $b$ of the metal may become comparable to $a/2$. As a result, the image potential may change. When $b < a/2$ one can think that the electric field produced by ions at the metal surface is relatively weak and slowly-varying. In such a case the screening charge of the metal is effectively situated at the distance $b$ away from the metal surface, \textit{i.e.} at $z=-b$. Therefore, the reflection plane for the image charge is at $z=-b$, so that an ion at distance $z$ from the surface experiences a smaller attraction $-e^2/4\varepsilon(z+b)$ to the surface, rather than the standard $-e^2/4\varepsilon z$ for a perfect metal. At the distance of closest approach $z=a/2$, the ion-to-surface attraction energy becomes $-e^2/2\varepsilon(a+2b)$. This leads to a finite energy cost $E_0$ for dissociating a bulk ion pair and bringing it to the metal surface, given by $E_0 = e^2/\varepsilon [a^{-1} - (a + 2b)^{-1}]^{-1}$. Thus, a finite voltage is necessary to break pairs in the bulk and obtain free ions which can provide screening. This means that, for an electrode with finite screening radius $b$, the bell-shaped $C(V)$ curve splits into two peaks located at $V=\pm E_0/e$, thereby becoming camel-shaped, in agreement with the above-mentioned data. 

The capacitance $C(V)$ at $V > E_0/e $ can be estimated with the help of the theory in section IV leading up to Eq.\ (\ref{eq:CV}). In this case, however, the voltage $V$ in Eq.\ (\ref{eq:CV}) should be replaced by $V-E_0/e$ and the dipole arm $a$ should be replaced by the longer arm $a+2b$. These substitutions give
\be 
C(V) = 1.4 \left( \frac{e}{\varepsilon (a +2b)(V-E_0/e)}
\right)^{1/3} \frac{a}{a+2b}C_H \label{eq:CV1} 
\ee
for $V > E_0/e$. Since the dipole-dipole repulsion is much stronger due to the longer dipole arm, $C(V)$ is substantially smaller and reaches its geometrical limit $C_H(b) = C_H \cdot a/(a+2b)$ at a smaller voltage $V-E_0/e = 5 \alpha e/4 \varepsilon (a+2b)$, or $V^* - E_0/(e^2/\varepsilon a) \approx 5.5/(1+2b/a)$.  Starting from this voltage the capacitance saturates at the level of $C_{H}(b)$.

Only at even larger voltage $V^* - E_0/(e^2/\varepsilon a) \gtrsim [5.5 a^2 + 8 \pi b(a+b)]/[a(a +2 b)]$ do counterions comprise a full layer at the surface, after which the theory of multi-layer arrangement of ions \cite{Kornyshev2007} becomes applicable.  This same behavior for $C(V)$ is expected in the case where a good metal is covered by a thin insulating layer, for example, its own oxide.

\begin{figure}[htb]
\centering
\includegraphics[width=0.45 \textwidth]{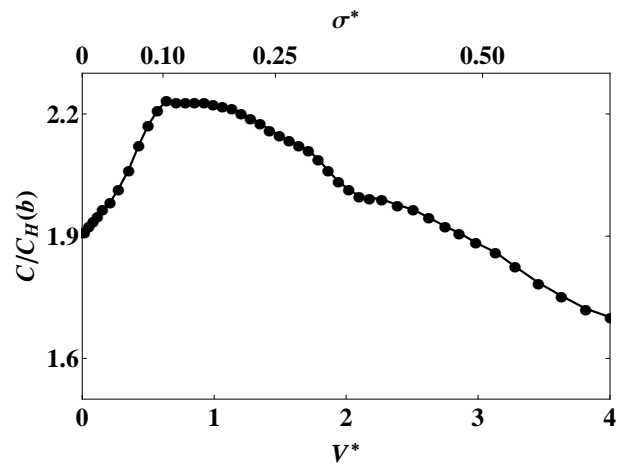}
\caption{The ratio of the capacitance of the metal/ionic liquid interface $C$ to the geometrical capacitance $C_H(b)$, plotted as a function of dimensionless voltage $V^*$ (bottom axis) and charge density $\sigma^*$ (top axis) for a system with ion density $Na^3=0.4$ and metal screening radius $b = a/2$. The capacitance is determined by a numerical derivative of the $\sigma$ vs.\ $V$ data obtained from a MC simulation at temperature $T^*=0.042$.  For $\varepsilon = 3$ and $a = 1$ nm, $V^* = 4$ corresponds to 1.92 Volts.} \label{fig:insulator}
\end{figure}

In order to verify this theory we repeated our MC calculations for $T^*=0.04$ and $Na^3=0.5$ using a relatively large $b=a/2$, which is at the limit of applicability for linear screening by the electrode surface. For simplicity, we have also assumed that the metal ion lattice has the same dielectric constant as our ionic liquid. Results are shown on Fig.\ \ref{fig:insulator}, plotted as a function of $V^*$ and $\sigma^*$. As expected, the peak at $V=0$ is split into a camel-like structure (we show only the positive half of the symmetric $C$--$V$ curve). The characteristic dimensionless voltage of the peak is $V^*_p \sim 0.5$, in agreement with the above estimate for $E_0$. Note that the capacitance maximum occurs at $\sigma^* = 0.1$ and is apparently not related to excluded volume effects among counterions.  As predicted by Eq.\ (\ref{eq:CV1}), the peak  capacitance $C(V^*_p)$ is approximately 2.5 times smaller than at $b = 0$ (recall that $C_H(b)$ in Fig.\ \ref{fig:insulator} is twice smaller than $C_H$ in Fig.\ \ref{fig:CV}).  

Figs.\ 1--3 clearly show that, for both a perfect metal $(b=0)$ and a semi-metal with $b=a/2$, the capacitance $C(V)$ can be 2--3 times larger than the geometrical mean-field capacitance $C_H(b)$ at moderately low temperature and voltage. One can interpret this fact by saying that the geometrical capacitance is in series with a negative capacitance from the ionic liquid, $C_{il}$, so that $C^{-1} =C_H(b)^{-1} + C_{il}^{-1}  < C_H(b)^{-1}$. Multiplying this equation by $\varepsilon/4\pi$ we arrive at $d^* = a/2 + b + r_s$, where $r_s$ is the screening radius of the ionic liquid. As we saw, at low temperatures and voltages this definition leads to negative $r_s$. At high temperatures $T^* > 1$, the screening radius is positive and given by Eq.\ (\ref{eq:rs}): $r_s = r_{DH}$. Then, as explained in the introduction, $d^* \simeq r_s$.

\section{Analogy with semiconductor devices}

Capacitance smaller than the geometrical one, or in other
words negative screening radius $r_s$, is well-known in semiconductor physics for capacitors made of a metal, an insulator of width $d$, and a semiconductor containing a clean two-dimensional electron gas (2DEG) with two-dimensional density $n$.  This can be, for example, a Si MOSFET or a gated GaAs-GaAlAs heterostructure.  In the limit of low density $n$, a 2DEG is a classical system whose physics is dominated by the Coulomb interaction between electrons, leading to a Wigner crystal-like strongly-correlated liquid state.  This state was shown to have a negative thermodynamic density of states and a negative screening radius  $r_s = - 0.116 n^{-1/2}$.  When $n^{-1/2}\ll d$, this screening radius produces a small negative correction to the geometric capacitance width $d$, so that  ~\cite{BLES1981,KRAV1990,Eis1992,Efros08}
\be 
d^* = d - 0.116 n^{-1/2}, \hspace{5mm} (n^{-1/2} \ll d). \label{eq:DSTAR}
\ee  

The question of what happens to $d^*$ when $n^{-1/2} \gg d$ has never been addressed.  Now we understand that in this limit one should think about the 2DEG and its image charges as a gas of electron-image dipoles oriented along the $z$ axis with dipole arms $2d$ (we consider the electrons, unlike ions, to be point-like particles so that for this case $a = 0$).  Repeating the calculations leading to Eq.\ (\ref{eq:CV}), again using the point-dipole approximation for the dipole-dipole repulsion, leads at low $T$ to an anomalously large capacitance $C(V)$ corresponding to the effective capacitor thickness 
\be 
d^* = 2.68 d \sqrt{n d^2} \ll d, \hspace{5mm} (n^{-1/2} \gg d) \label{eq:DSTAR1}.
\ee  

\begin{figure}[htb]
\centering
\includegraphics[width=0.45 \textwidth]{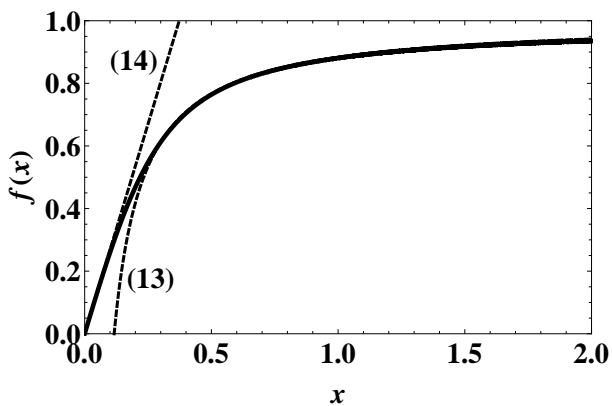}
\caption{The scaling function $f(x)$ which determines the effective thickness $d^*$ of a capacitor composed of a 2DEG with area density $n$ separated from a perfect metal surface by an insulating layer of thickness $d$, as defined by Eq.\ (\ref{eq:f}).  The left side of the plot corresponds to a very sparse 2DEG, where the electrons can be thought to form an array of discrete ion-image dipoles and $d^*$ is described by Eq.\ (\ref{eq:DSTAR1}).  The right side corresponds to a relatively dense packing of electrons, where the electrons approach a uniform layer of charge and $d^*$ is described by Eq.\ (\ref{eq:DSTAR}).} \label{fig:numericsum}
\end{figure}

In order to study the crossover between Eq.\ (\ref{eq:DSTAR}) and Eq.\ (\ref{eq:DSTAR1}), we calculated a universal low temperature function $d^*(d, n)$ through an exact numerical calculation of the interaction energy of an infinite lattice of dipoles with dipole arm $2d$ and areal density $n$, abandoning the point dipole approximation.  The result can be written as
\be 
d^* = d \cdot f(d n^{1/2}), \label{eq:f}
\ee
where $f(x)$ is a dimensionless function shown in Fig.\ \ref{fig:numericsum}. In the case of a low-temperature ionic liquid with diameter small enough that $na^2 \ll 1$, one can use this result by replacing $d$ with $d + a/2$. Then Eq.\ (\ref{eq:f}) describes the shape of the crossover of $C(V)$ to the geometrical capacitance $C_H(b)$.

We note that the predictions of this section might be verifiable in an extremely clean GaAs-GaAlAs heterojunction with an area density of holes lower than $n=10^9$ cm$^{-2}$~\cite{Huang}, assuming that one is able to make gates even closer than the current limit of $250$ nm without an increase in disorder.  Our theory might also be verified in a capacitor composed of a 2DEG on top of a thin film of liquid Helium covering a metal gate~\cite{Grimes}.

\
\section{Conclusion}

To summarize, this paper is concerned with the restricted
primitive model of an ionic liquid at a metal interface. Within
this model, we obtain capacitance at zero voltage as large as $3C_H$.  We also find that for a good metallic electrode at small voltage, $C$ decreases with $T$ as $T^{-1/3}$.  When the temperature is fixed and is relatively small, $C(V)$ decreases as $1/V^{1/3}$ (the $C$--$V$ curve is ``bell-shaped"). On the other hand, when the electrode is made from a semi-metal the $C$--$V$ curve is ``camel-shaped", meaning that the capacitance first grows with $V$ and then goes through a maximum and decays as $1/V^{1/3}$. We interpret these results with the help of a semi-quantitative analytical theory based on the weak repulsion between ion-image dipoles, and we confirm our results with a MC simulation. Our conclusions are in qualitative agreement with experimental data.

We are grateful to D. Boda, A. Kornyshev, Y. Levin, C. Outhwaite, P. A. Madden, and M. B. Partenskii for helpful discussions.



\begin{thebibliography}{100}
\bibitem{clusters} 
This statement is correct even in a relatively dense liquid, where
a significant fraction of ion pairs in the bulk form larger neutral clusters, and ions sticking to the metal may be decorated by one or more neutral pairs. Let us imagine a bulk cluster made of $2m$ ions, where $m$ is an odd number --- for example, $m=3$.  If the cluster is sufficiently symmetric, then it can be dissociated into two smaller clusters, one positive and one negative, which have an equal number $m$ of ions and which differ only by the inversion of the charge of each ion.  When these two charged clusters stick to their images in the metal surface they also recover the energy spent on their dissociation.  Thus we are dealing with a statistical physics problem of very high degeneracy, which may have a beautiful exact solution. (Less symmetric clusters may have a small contribution to the partition function of the system.) In this paper we use the term ``excess ions" to refer to all charged objects at the surface, regardless of whether they are decorated by neutral pairs.
\end{thebibliography}
\end{document}